\documentclass[amsmath,amssymb,eqsecnum,superscriptaddress,prx,twocolumn,showpacs]{revtex4}
\usepackage{graphicx}
\usepackage{dcolumn}
\usepackage{bm}
\usepackage{mathrsfs}
\usepackage{amssymb}
\usepackage{amsmath}
\usepackage{amsfonts}
\usepackage{braket}
\usepackage{epsfig}
\usepackage{natbib}
\usepackage{mathdots}
\usepackage{multirow}
\usepackage{epstopdf}
\usepackage{hyperref}
\usepackage{float}	
\usepackage{bibentry}
\usepackage[caption=false]{subfig}
\newcommand{\ba}{\begin{eqnarray}}
\newcommand{\ea}{\end{eqnarray}}
\newcommand{\bd}{\begin{displaymath}}
\newcommand{\ed}{\end{displaymath}}
\renewcommand{\v}[1]{{\bf #1}}
\newcommand{\bpm}{\begin{pmatrix}}
\newcommand{\epm}{\end{pmatrix}}
\newcommand{\nn}{\nonumber \\}

\graphicspath{{figures/}}

\begin{document}

\title{Proposal of a spin-one chain model with competing dimer and trimer interactions}
\author{Yun-Tak Oh}
\affiliation{Department of Physics, Sungkyunkwan University, Suwon 16419, Korea}
\author{Hosho Katsura}
\email[Electronic address:$~~$]{katsura@phys.s.u-tokyo.ac.jp}
\affiliation{Department of Physics, Graduate School of Science, The University of Tokyo, Hongo, Bunkyo-ku, Tokyo 113-0033, Japan}
\author{Hyun-Yong Lee}
\affiliation{Institute for Solid State Physics, University of Tokyo, Kashiwa, Chiba 277-8581, Japan}
\author{Jung Hoon Han}
\email[Electronic address:$~~$]{hanjh@skku.edu}
\affiliation{Department of Physics, Sungkyunkwan University, Suwon 16419, Korea}
\date{\today}

\begin{abstract}
A new kind of spin-1 chain Hamiltonian consisting of competing dimer and trimer projection operators is proposed. As the relative strengths and signs of the interactions are varied, the model exhibits a number of different phases including the gapped dimer phase and the gapless trimer phase with critical correlations described by a conformal field theory with central charge $c=2$. A symmetry-protected topological phase also exists in this model, even though the microscopic interactions are not the simple adiabatic extensions of the well-known Heisenberg and the Affleck--Kennedy--Lieb--Tasaki model and contains both two- and three-particle permutations. A fourth phase is characterized by macroscopically degenerate ground states. While bearing almost a one-to-one resemblance to the phase diagram of the bilinear-biquadratic spin-1 chain Hamiltonian, our model is rooted on very different physical origin, namely the two competing tendencies of spin-1 particles to form singlets through either dimer or trimer formation.
\end{abstract}
\maketitle

\section{Introduction}
Resonating valence bond (RVB) state has been studied for decades since Anderson proposed it as the ground state of antiferromagnetic spin-$1/2$ Heisenberg model on the triangular lattice \cite{anderson73}. After the Rokhsar-Kivelson proposal for a short-range dimer RVB on the square lattice \cite{rokhsar88} followed by the Moessner-Sondhi proposal to the effect on the triangular lattice\cite{moessner01}, quantum dimer model and the dimer RVB state have been studied thoroughly for various lattices. A good review of our understanding of the short-range dimer RVB can be found in Ref.  \onlinecite{moessner11}. As a general consensus, a dimer liquid phase with $\mathbb{Z}_2$ topological order forms naturally in the non-bipartite lattice, but not so in the bipartite lattice.

Recently, we proposed the trimer version of RVB state on the square lattice \cite{lee17}. It was carefully argued that a liquid phase with $\mathbb{Z}_3$ topological order should be realized by the trimer model despite the bipartite nature of the square lattice. Trimers assumed in Ref. \onlinecite{lee17} reflect the spin singlet made from three spin-1 particles in an appropriately chosen spin Hamiltonian. A natural but challenging question is whether one can write down some microscopic spin-1 Hamiltonian supporting the trimer spin liquid phase. To be clear, we refer to the spin singlet formed by three adjacent spin-1 objects as the trimer.

In quantum trimer models such as proposed by the present authors~\cite{lee17}, as well as in all quantum dimer models~\cite{moessner11} or the recently proposed dimer-pentamer model~\cite{myers17}, orthogonality of different dimer, trimer, or dimer-pentamer configurations are assumed from the outset. Treated as a real spin singlet, of course, the orthogonality is lost due to the singlet breaking into higher-spin configurations, and this is the main reason that writing down an exact microscopic spin Hamiltonian for resonating dimers becomes very hard \cite{fujimoto05,seidel09,cano10}.
We can instead work with a simple enough spin Hamiltonian that embodies a state quite like the quantum dimer or trimer liquid. Due to the general challenge in writing down two-dimensional spin models and in solving them reliably, we are more likely to address the question effectively in the one-dimensional context first.

\begin{figure}[htbp]
\includegraphics[width=0.48\textwidth]{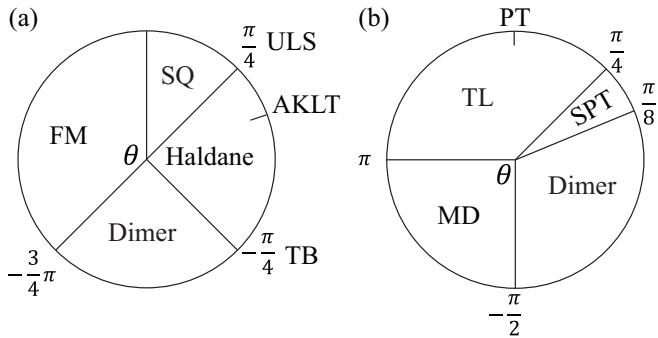}
\caption{Phase diagram of (a) BLBQ model and (b) DT model as a function of the mixing angle $\theta$. Abbreviations stand for FM$=$ferromagnetic,  MD$=$macroscopically degenerate, SPT $=$ symmetry-protected topological, SQ$=$spin-quadrupolar, TL$=$trimer liquid. In both phase diagrams, the dimer phase is topologically trivial and gapped, with ground states that break the translation symmetry. Haldane (SPT) phase is topologically non-trivial and translation-invariant. SQ and trimer phases are both critical and carry the central charge $c=2$. MD phase has exponentially large ground state degeneracy. The pure trimer (PT) point on the right phase diagram possesses the full SU(3) symmetry, as does the ULS model on the left.  }
\label{fig:PD}
\end{figure}

Trimerization in spin-1 chain model has had an interesting history. For instance, the issue drew significant attention in the context of spin-1 bilinear-biquadratic (BLBQ) Hamiltonian

\begin{align}
H_{\rm BLBQ} = \sum_{i=1}^N \Bigl[ \cos \theta ( \v S_i \cdot \v S_{i+1} ) + \sin \theta (\v S_i \cdot \v S_{i+1})^2 \Bigr] . \label{eq:BLBQ-H}
\end{align}
This model encompasses the spin-1 antiferromagnetic Heisenberg model at $\theta=0$,~\cite{haldane83} exactly solvable Affleck-Kennedy-Lieb-Tasaki (AKLT) Hamiltonian at $\tan \theta=\frac{1}{3}$,~\cite{AKLT87,AKLT88} and several integrable models~\cite{sutherland75,takhtajan82,babujian82,parkinson88,barber89,klumper89,klumper90_1,klumper90_2} as special points. The phase diagram of this model Hamiltonian has been carved out over the past several decades (see Fig. \ref{fig:PD})\cite{sutherland75,takhtajan82,babujian82,parkinson88,barber89,klumper89,klumper90_1,klumper90_2,nomura91,fath91,xian93,itoi97,schmitt98,lauchli06}.  Undoubtedly the most important phase of the BLBQ model is the Haldane phase, realized over $-\pi/4 < \theta < \pi/4$, which is a translationally invariant state with massive excitations. More recently, the Haldane phase came to be identified as an example of the symmetry-protected topological phase, or SPT for short, with an intriguing double degeneracy in the entanglement spectrum protected by discrete symmetries~\cite{pollmann10,chen11}.

Technically, the phase that took a considerable amount of effort in clarifying its nature exists over the $\pi/4 < \theta < \pi/2$ region of the BLBQ model. The $\theta=\pi/4$ point is the well-known Uimin-Lai-Sutherland (ULS) Hamiltonian\cite{uimin70,lai74,sutherland75}, which has the enhanced SU(3) symmetry despite being a spin-1 model. It is solvable by Bethe ansatz and possesses gapless spinon modes. An early pioneering numerical study by F\'{a}th and S\'{o}lyom has seen signatures of period-3 oscillations in various physical observables.
Possibilities of the trimerized ground state have been raised by several theorists \cite{xian93,nomura91}. An exact Hamiltonian for the three-fold degenerate trimer solid ground state was proposed by Schmitt {\it et al}. \cite{schmitt98}, S\'{o}lyom and Zittartz \cite{solyom00}, and more recently by Rachel and Greiter~\cite{rachel08}, in a generalization of the Majumdar-Ghosh Hamiltonian of the dimer solid ground state of spin-$1/2$ chain \cite{majumdar69}. As the exact model construction for the trimer leaned in favor of the gapped ground state with translational symmetry breaking, the numerics of F\'{a}th and S\'{o}lyom has come down on the side of gapless phase for $\pi/4 < \theta < \pi/2$ region of the BLBQ model. These days, this region is best described as the spin quadrupolar (SQ) phase after the numerical works such as Refs. \onlinecite{schmitt98,lauchli06} that tried to identify the dominant correlations in this phase.

In this paper, we propose a new class of spin-1 Hamiltonians, motivated by the simple observation that there are two ways in which $S=1$ spins can form a singlet: one is by dimerizing the two adjacent spins, and the other is by trimerizing the three adjacent spins. Singlet formation over more than three sites is neglected. We thus consider a model that consists of dimer and trimer projection operators as
\begin{align}
H_{\rm DT} = -\sum_{i}\bigl( \cos \theta \, D(i) + \sin \theta \,  T(i) \bigr).
\label{eq:DT_model}\end{align}
This will be called the dimer-trimer (DT) Hamiltonian throughout the paper. The operators $D(i)$ and $T(i)$ are proportional to the dimer and trimer projection operators, respectively, to be defined precisely in the next section.

The rest of the paper concerns the analysis of the proposed DT Hamiltonian. Dimer and trimer projection operators are introduced in Sec. \ref{sec:DT_operators}. In Sec. \ref{sec:phase_diagram}, the phase diagram of the DT model is worked out as a function of $\theta$ using the powerful density-matrix renormalization group (DMRG) method of identifying the ground state. Four phases are identified: dimer, SPT, trimer liquid, and macroscopically degenerate, respectively. The dimer phase is gapped and breaks the translational symmetry of the lattice.
The trimer liquid phase is critical and shares many physical properties with the spin quadrupolar phase of the BLBQ Hamiltonian. The SPT phase exhibits the even-number degeneracy in the entanglement spectrum that remains robust against perturbations~\cite{pollmann10}. The macroscopically degenerate phase literally carries the ground state degeneracy that grows exponentially with the lattice size. Interesting parallel with, and differences from, the phase diagram of the BLBQ model Hamiltonian is pointed out along the way.

\section{dimer and trimer projection operators}
\label{sec:DT_operators}

Our first task is to give proper definition to dimer and trimer operators. Using the spin-1 operator $\v S_i$ at each lattice site, $\v S_{ij} = \v S_i + \v S_j$ for a pair of adjacent sites ($j=i+1$), and $\v S_{ijk} = \v S_i + \v S_j+ \v S_k$ for a triplet of adjacent sites ($k=i+2$), the dimer and trimer projection operators can be expressed as
\begin{align}
{\mathcal P}_D(i) & =  \frac{1}{12} \left( \v S_{ij}^2 -2 \right) \left(\v  S_{ij}^2 -6\right)=  \frac{1}{3} \left( \v S_i \cdot \v S_{j}\right)^2 -\frac{1}{3} , \nn
{\mathcal P}_T(i) & =  - \frac{1}{144}  \left(\v S_{ijk}^2 - 2\right) \left(\v S_{ijk}^2 - 6 \right) \left(\v S_{ijk}^2 - 12\right). \label{eq:dimer-and-trimer-projector}
\end{align}
Each projection operator gives $+1$ for the spin singlets, and $0$ for all other spin multiplets.  The projectors are related to $D(i)$ and $T(i)$ in the DT Hamiltonian (\ref{eq:DT_model}) by
\begin{align}
D(i) =  3 {\mathcal P}_D(i) , ~~~
T(i) = 6 {\mathcal P}_T(i) .
\label{eq:DT_operators}\end{align}

It is not widely appreciated in the literature, and therefore requires some highlighting here, that the dimer projection operator is equivalent to the pure-biquadratic (PBQ) spin interaction up to constants. The PBQ model, corresponding to $\theta=0$ in the DT Hamiltonian, is known to be exactly solvable and possess a hidden SU(3) symmetry~\cite{parkinson88,barber89,klumper89,klumper90_1,klumper90_2,affleck90}.

The trimer projection operator looks much more complicated by contrast, and involves three-site spin interactions. The trimer singlet state given by the totally anti-symmetric combination
\begin{align}
|{\rm trimer} \rangle = \frac{1}{\sqrt{6}}\sum_{a,b,c} \varepsilon_{abc} |a, b, c\rangle , \label{eq:trimer-wf}
\end{align}
where $a,b,c = +1, 0, -1$ refers to the $S^z$ eigenvalue, is invariant under the SU(3) transformation, and $\varepsilon_{abc}$ is the antisymmetric tensor. This observation prompted us to seek an alternative expression in terms of Gell-Mann matrices, and subsequently use the identity
relating the inner product of Gell-Mann matrices to the exchange operator $P_{ij}$:
%
%
\begin{equation}
\v \Lambda_i \cdot \v \Lambda_j  = 2 P_{ij} -\frac{2}{3} ,
\label{eq:Gell-Mann_exchange}
\end{equation}
where $\v \Lambda_i = (\Lambda^1_i , \cdots , \Lambda^8_i )$ is s a collection of eight Gell-Mann operators at site $i$.
The exchange operator $P_{ij}$ swaps the state at site $i$ with the state at site $j$.
%
%
By using the exchange operators, we can arrive at an interesting alternative expression of the trimer projector:
\begin{align}
{\mathcal P}_T(i) = \frac{1}{6} \Bigl( 1+ P_{ijk} + P^{-1}_{ijk} - P_{ij} -  P_{jk} - P_{ki} \Bigr) .\label{eq:permutation}
\end{align}
The three-site ring exchange operators \cite{thouless65} are introduced as
\begin{align}\label{eq:ring_exchange}
P_{ijk} = &  P_{jk}P_{ij} = P_{ik}P_{jk} = P_{ij}P_{ik},\nn
P_{ijk}^{-1} = & P_{ij}P_{jk} = P_{jk}P_{ik} = P_{ik}P_{ij} .
\end{align}
The trimer projector is a sum of three-spin exchange among the three adjacent sites, minus the pairwise exchange for adjacent and second-adjacent sites. Recalling the relation $\v S_i \cdot \v S_{j}+\left( \v S_i \cdot \v S_{j}\right)^2 = P_{ij} +1$, the dimer projection operator can be expressed as
${\mathcal P}_D(i) = \frac{1}{3} \left( P_{ij} - \v S_i \cdot \v S_{j} \right)$.

The DT model coincides with the BLBQ Hamiltonian at the two points, $\theta=0,\pi$. Otherwise, the nature of the ground states for other values of $\theta$ remains to be explored. We have carried out the DMRG calculation, keeping 3000 Schmidt states during the iteration and employing open boundary condition on a chain of length $L=90$, to work out the phase diagram of the DT model.
ITensor library \cite{itensor} is employed in the single-site DMRG calculation with the noise algorithm \cite{white05}. By varying $\theta$, we can carve out most of the phase diagram for $-\pi/2 \le \theta \le \pi$. Ground states for $\pi \le \theta \le 3\pi/2$ are hard to reach by DMRG due to the large number of ground state degeneracies. We use analytic arguments and exact diagonalization study of the Hamiltonian to gain understanding of the ground states here.

\section{Phase Diagram}
\label{sec:phase_diagram}

Several order parameters and their correlations were calculated for each $\theta$, in addition to the spin-spin correlation function. They include the dimer average $\langle {\cal P}_D (n) \rangle$, the trimer average $\langle {\cal P}_T (n) \rangle$, and their connected correlation functions, $\langle {\cal P}_D (x){\cal P}_D (x+n)\rangle-\langle {\cal P}_D (x)\rangle\langle{\cal P}_D (x+n)\rangle$ and $\langle {\cal P}_T (x){\cal P}_T (x+n)\rangle-\langle {\cal P}_T (x)\rangle\langle{\cal P}_T (x+n)\rangle$. Here, we set $x=N/4$, which is sufficient depth that the edge effects died out. Note that we are using the dimer projector ${\cal P}_D$ as a measure of the dimer order rather than $\v S_i \cdot \v S_j$, which is the more commonly used measure of dimer correlations in the literature. Results using both order parameters lead to similar conclusions.

Figure \ref{fig:PD} shows the phase diagram of the DT model alongside the well-known one for the BLBQ model. We were able to identify four phases altogether, bearing a close resemblance to the BLBQ model despite very different character of microscopic interactions in the two models.

\begin{figure}[htbp]
\includegraphics*[width=0.485\textwidth]{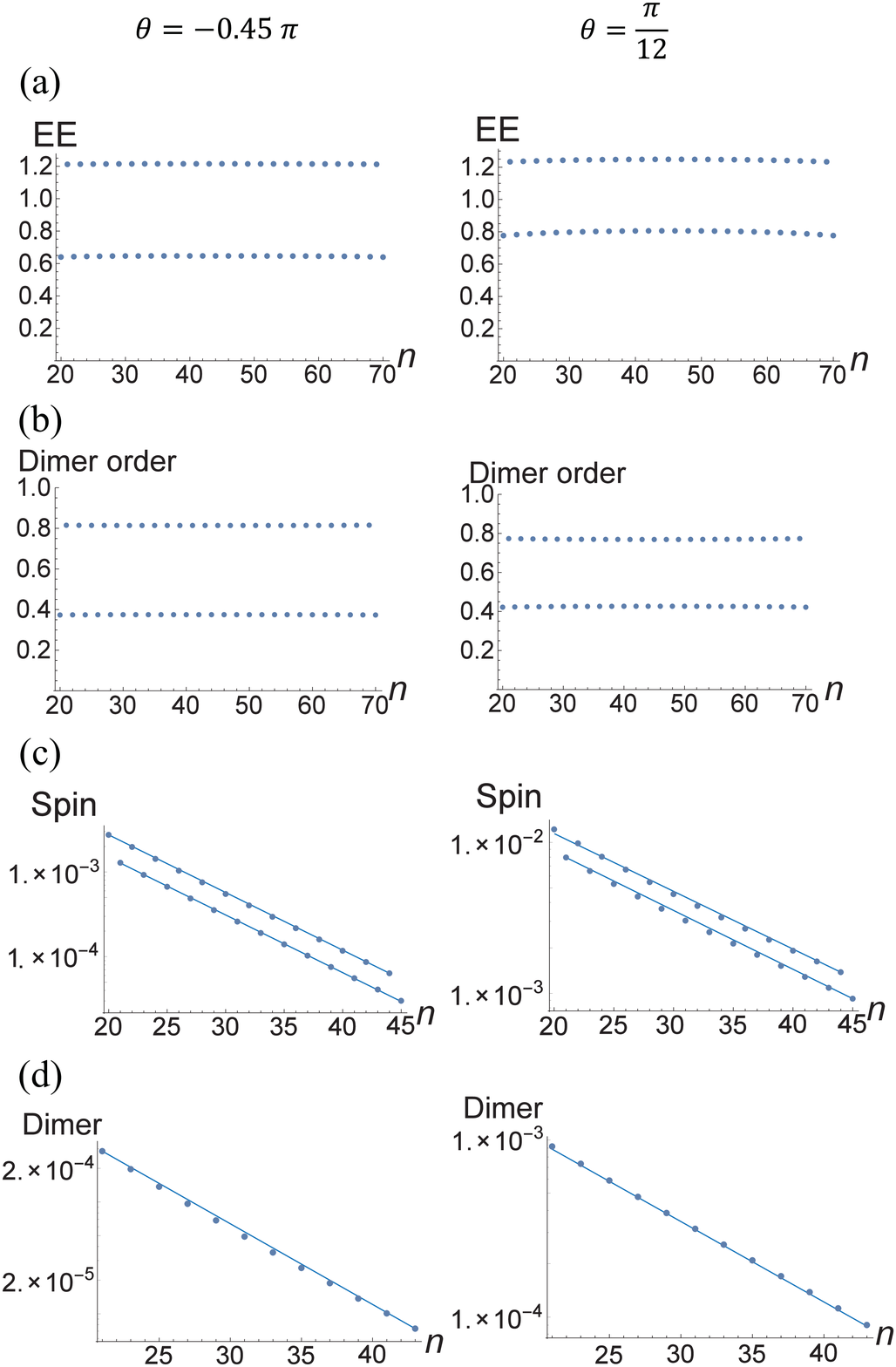}
\caption{(a) Entanglement entropy, (b) dimer density $\langle {\mathcal P}_D(n)\rangle$, (c) spin-spin correlation, and (d) dimer-dimer correlation for $\theta=-0.45\pi$ (left column) close to the phase boundary and $\theta=\pi/12$ (right column) deep inside the dimer phase. Period-2 oscillations appear in all the calculated quantities. Correlation functions in log-linear plots (c) and (d) decay exponentially.}
\label{fig:dimer}
\end{figure}

\subsection{Dimer phase}

It is known that the $\theta=0$, pure biquadratic Hamiltonian has the dimerized ground state with $\langle {\cal P}_D (n) \rangle_{\rm PBQ} \approx D_0 + D_1 (-1)^n$, in reflection of the spontaneous translational symmetry breaking in the ground state\cite{barber89,klumper89,klumper90_1,klumper90_2}. Numerical data shown in Fig. \ref{fig:dimer} indeed finds such binary oscillations of the physical observables in space, as well as in the entanglement entropy (EE).  The trimer average remains zero within the resolution of our calculation.

Evidence of the gap in the energy spectrum comes from exponentially decaying correlation functions shown in Fig. \ref{fig:dimer} (c)-(d).
%
%
Starting from the well-known dimer state at $\theta=0$, we conclude that the extraneous trimer interaction coming from $\theta \neq 0$ is not destroying the dimer order thanks to the protection from the energy gap. The whole region $-\pi/2 < \theta \lesssim \pi/8$ shows nearly identical behavior in the physical quantities calculated, bolstering the claim that this is the same dimer phase as seen at the pure-biquadratic point. As a final indicator of the dimer phase, the lowest level in the entanglement spectrum shows the characteristic alternation between single and triple degeneracy depending on the even and odd position of the cut. The same feature exists in the dimer phase of the BLBQ Hamiltonian~\cite{thomale15}.

\subsection{SPT phase}

\begin{figure}[htbp]
\includegraphics*[width=0.48\textwidth]{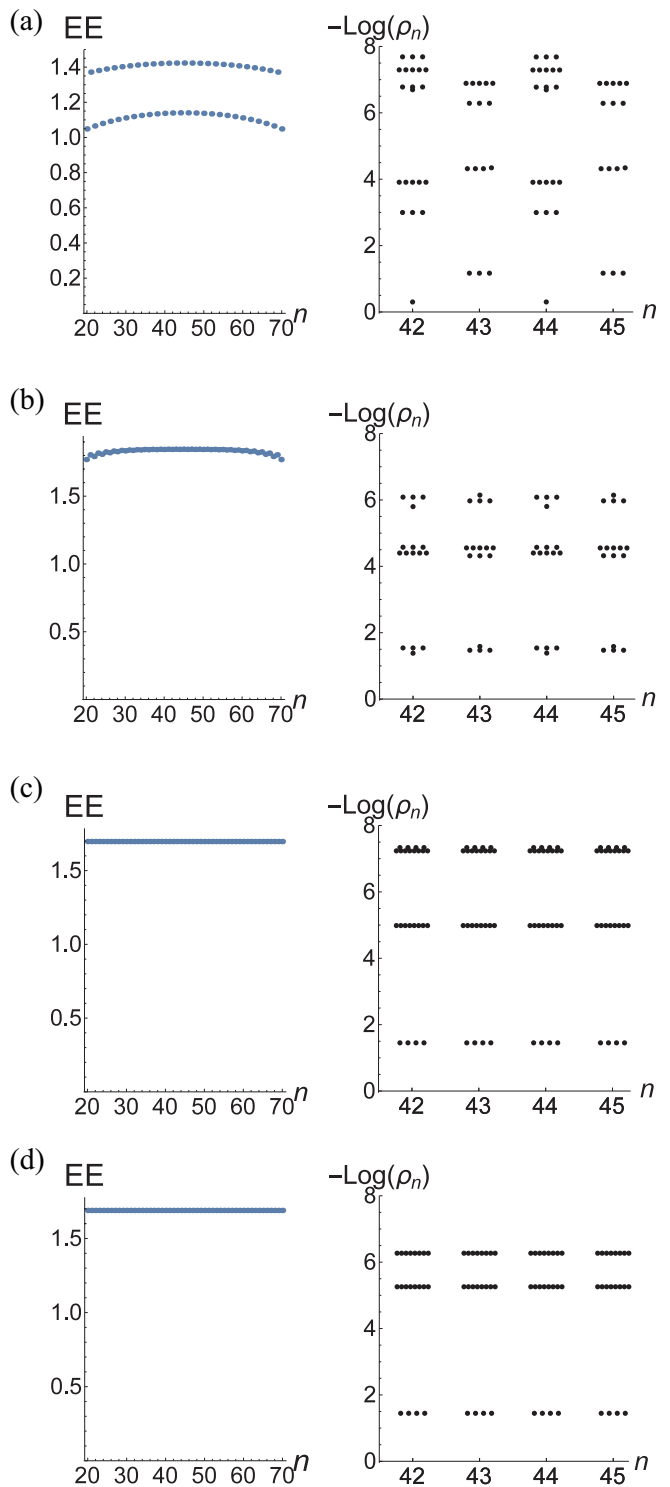}
\caption{Entanglement entropy (left column) and entanglement spectrum near the middle of the lattice (right column) in the dimer phase (a) $\theta=\pi/8$, and in the SPT phase (b) $13 \pi / 96$, (c) $14 \pi / 96$, and (d) $15 \pi / 96$. At $\theta =  \pi/8$, EE still has the period-2 oscillation and ES shows the alternating $3-1-\cdots$ degeneracy of the lowest levels characteristic of the dimer phase. At $\theta = 13 \pi/96$, EE loses the period-2 structure and ES shows the 4-fold (sometimes 8-fold or 12-fold) degeneracy in the entire spectra. The trend continues for $\theta=14 \pi/96$ and $15 \pi/96$.  }
\label{fig:Ent_Haldane}
\end{figure}

A very different feature emerges as soon as one moves past $\theta=\pi/8$. While all the calculated quantities for $\theta=\pi/8 = 12\pi/96$ have the characteristic properties of the dimer phase - {\it e.g.} compare Fig. \ref{fig:dimer}(a) to Fig. \ref{fig:Ent_Haldane}(a) - the same calculation done at $\theta=13\pi/96$ gives a completely different picture; see Fig. \ref{fig:Ent_Haldane}(b)-(d). Even with our choice of finely spaced angles $\theta$, it was hard to detect gradual changes in calculated properties near the phase boundary, suggesting that it is likely the first-order phase transition separating the dimer from the other phase taking place for $\theta \gtrsim \pi/8$, which we came to identify with the SPT (Haldane) phase~\footnote{Our numerical investigation focuses on identifying the various phases of the DT model through calculations of physical and entanglement quantities. An unambiguous identification of the nature of the phase boundaries is delegated to another paper. At this point, we conclude that a continuous phase transition is not ruled out, but very unlikely.}

The period-2 oscillation vanishes completely in this phase in restoration of the translation symmetry. At the same time the trimer average becomes non-zero, $\langle {\cal P}_T (i) \rangle \neq 0$, while the dimer average continues to remain finite. A check on the Haldane phase of the BLBQ Hamiltonian confirms that these order parameters are nonzero there, too. It is still a gapped phase, as one can deduce from various exponential correlations and the flatness of the entanglement entropy shown in Fig. \ref{fig:Ent_Haldane}.

The most important check on the nature of the new phase comes from the entanglement spectrum. As in the Haldane phase of the BLBQ model \cite{pollmann10}, the entire entanglement spectrum of the DT model over this  phase has the degeneracy in multiples of four. For instance in Fig. \ref{fig:Ent_Haldane}, the fourfold degeneracy is only slightly imperfect at $\theta=13\pi/96$, presumably due to a finite-size effect, but completely restored at larger $\theta$. The Haldane phase of the BLBQ Hamiltonian has the characteristic degeneracy in multiples of two~\cite{pollmann10}.

\begin{figure}[htbp]
\includegraphics*[width=0.48\textwidth]{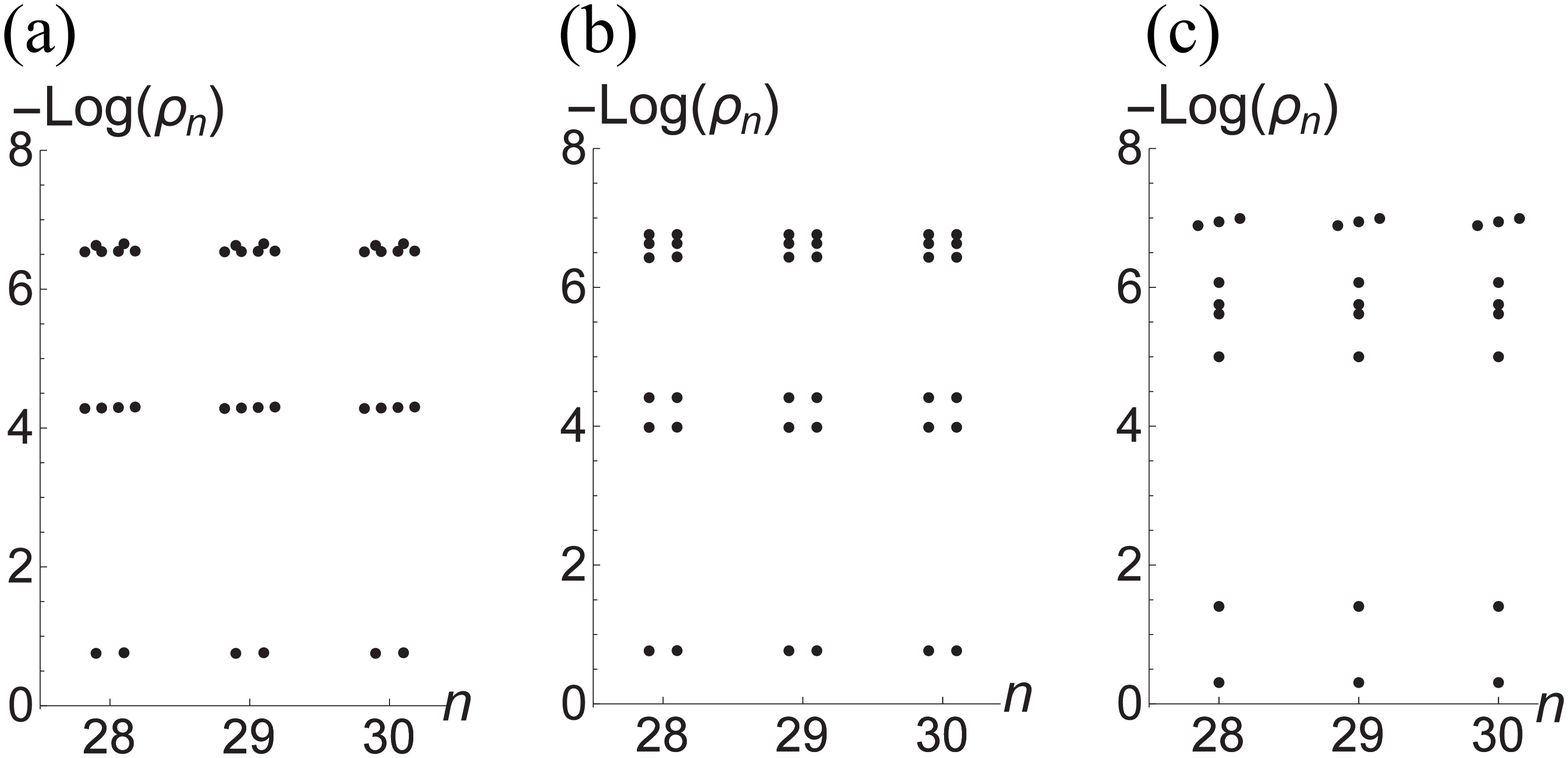}
\caption{Entanglement spectrum in the SPT phase at $\theta =  14 \pi/ 96$ and the system size $N=60$ for the DT model with extra perturbation (\ref{eq:delta-H}). (a) only $B\neq 0$. (b) Only $A$ and $B$ nonzero. Twofold degeneracy survives throughout the whole spectrum in both cases.  (c) only $C$ nonzero. The even degeneracy is lifted, signaling the destruction of SPT phase.}
\label{fig:Haldane_ES_sym}
\end{figure}

To better understand the nature of the degeneracy in the entanglement spectrum, we subject the DT Hamiltonian to various perturbations as suggested in Ref. \onlinecite{pollmann10}:
\begin{align}
\delta H = & A\sum_i (S_i^x S_i^y + S_i^y S_i^x )  + B \sum_i S^z_i \nn
& +  C \sum_i (S_i^z -S^z_{i+1}) ( S_i^x S_{i+1}^x + S_i^y S_{i+1}^y )  \nn
& +  C \sum_i (S_i^x S_{i+1}^x + S_i^y S_{i+1}^y )  (S_i^z -S_{i+1}^z )  . \label{eq:delta-H}
\end{align}
According to the SPT criterion of Ref. \onlinecite{pollmann10}, the addition of moderate amounts of $A$ and $B$ terms in the above should not break the even multiplicity of the entanglement spectrum, but the addition of the $C$ term should. As shown in Fig. \ref{fig:Haldane_ES_sym}, this is exactly what we find for the DT Hamiltonian in the putative topological phase. Although the fourfold degeneracy is easily lifted by finite $A$ and $B$, the twofold degeneracy survives the $A$ and $B$ perturbation as long as the symmetry-violating $C$-term is added.

After these stringent checks, we conclude that the region $\pi/8\lesssim\theta<\pi/4$ is a gapped topological phase which is commonly known as the SPT lately~\cite{pollmann10,chen11}.
For this range of $\theta$ values in the DT model, the two-site permutation, three-site ring exchange, and Heisenberg-type spin interaction all contribute. Although in its appearance this model does not resemble the AKLT or the Heisenberg Hamiltonian at all, we claim that they all belong to the same SPT phase.

A side remark is in order regarding the possible quantum numbers of the fourfold and the twofold degenerate entanglement levels in this phase. It is possible that the fourfold degenerate levels form an effective spin $S=3/2$ multiplet, or a pair of $S=1/2$ multiplets whose levels were ``accidentally" degenerate and immediately lifted by the addition of $A$ and $B$ terms. Unfortunately, it was not possible to compute the quantum numbers of the Schmidt states explicitly using our current numerical implementation of DMRG.  In either case, however, we may claim that spin quantum number fractionalization as $S=1/2$ or $S=3/2$ is taking place in an integer spin-chain model, in a reflection of the topological nature of this phase.

\begin{figure}[htbp]
\includegraphics*[width=0.45\textwidth]{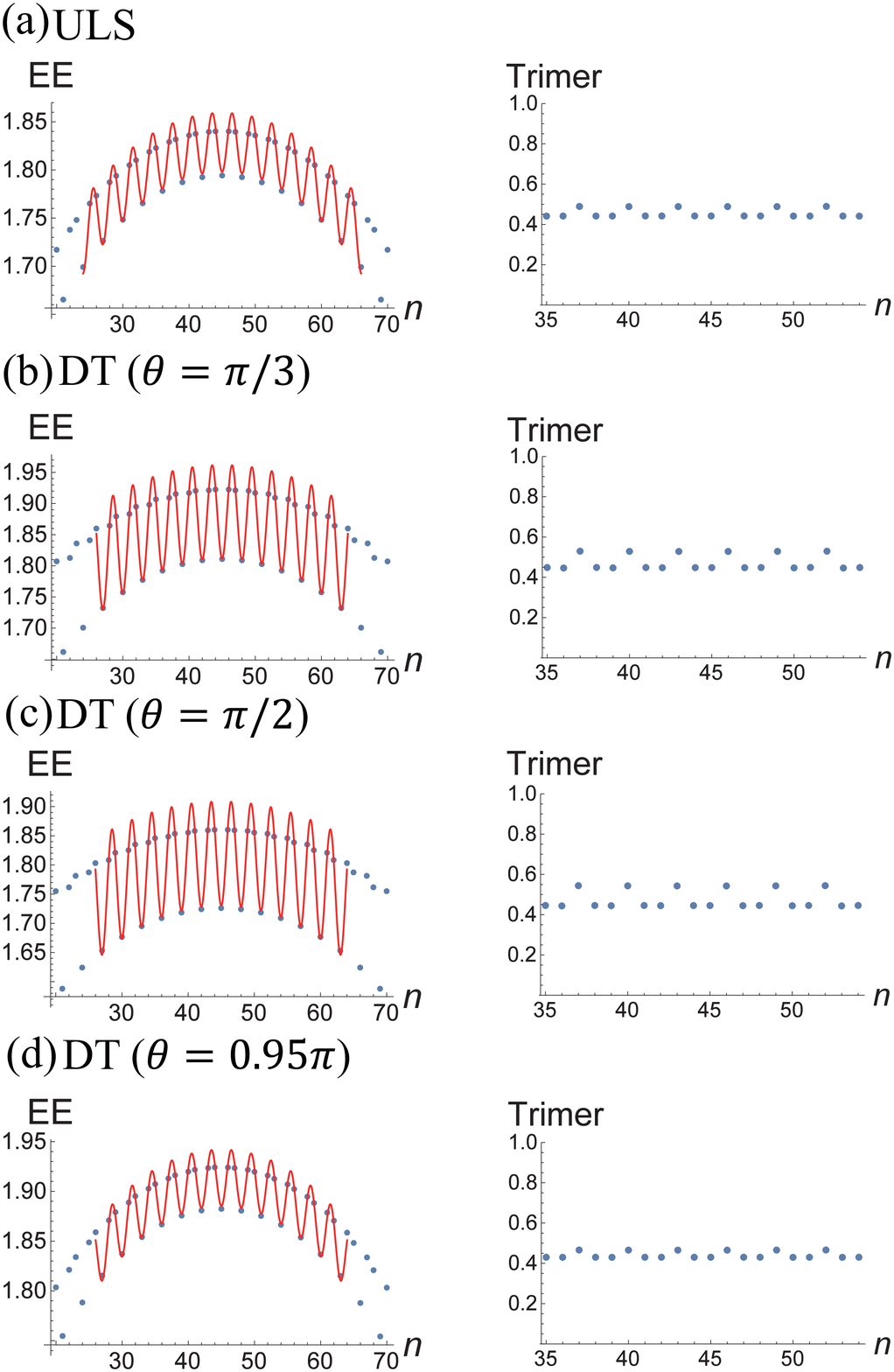}
\caption{Entanglement entropy (left column) and trimer average $\langle {\cal }P_T (n) \rangle$ for the trimer liquid phase at several $\theta$ values. The ULS results are shown in (a) for comparison \cite{PJkim16}. The red curves in the entanglement entropy plots indicate the fitting to the Calabrese-Cardy formula.}
\label{fig:trimer_EE_orders}
\end{figure}

\subsection{Trimer liquid phase}
Another phase appears as $\theta=\pi/4$ is crossed. As with the dimer-SPT phase boundary, our numerics in the SPT phase was not able to identify explicitly a gradual decrease in the energy gap or the divergence of correlation length near the phase boundary.
In the BLBQ Hamiltonian, the phase boundary separating the Haldane phase from the gapless, spin-quadrupolar phase was the ULS point in possession of exact SU(3) symmetry. The symmetry of the DT model at or around $\theta=\pi/4$ is {\it not} SU(3). The true SU(3)-symmetric point in the DT model is at $\theta=\pi/2$, lying deep inside this new phase. Based on a number of checks, we conclude the gapless phase is very similar in physical properties to the spin-quadrupolar phase in the BLBQ model realized over $\pi/4 < \theta < \pi$.
To be more precise, the phase is critical, with a good fit of the entanglement entropy behavior to the conformal field theory prediction with the central charge equal to $2$. The same statement applies to the spin-quadrupolar phase as well~\cite{lauchli06,thomale15}.

As shown in Fig. \ref{fig:trimer_EE_orders}, period-$3$ oscillations appear in the entanglement entropy and the trimer density $\langle {\cal P}_T \rangle$. The dimer order remains very close to zero throughout the trimer phase. The cut-position dependence of the entanglement entropy can be fit to the Calabrese-Cardy formula
\cite{calabrese04,D'Emidio15,PJkim16}
\begin{align}
S_n & = S_n^{\rm CFT} + S_n^{\rm osc} + c' \nn
S_n^{\rm CFT} & = \frac{c_N}{6} \log \left[\frac{2L}{\pi} \sin \left(\frac{\pi n}{L} \right) \right] \nn
S_n^{\rm osc} & = \sum_a F^a \left( \frac{n}{L} \right)\frac{\cos( 2 a \pi n/N)}{|L \sin(\pi n/L ) |^{\Delta_a}}.
\end{align}
Here, $S_n$ is the EE of the subsystem of length $n$, $c_N=N-1$ and $\Delta_a$ are the central charge and scaling dimension of the SU$(N)_1$ Wess--Zumino--Witten theory, respectively.  Non-universal constant $c'$ and the universal scaling function $F^a$, also adequately approximated as a constant, can be chosen to fit the EE data as well as one can. All the entanglement entropies calculated within the trimer phase fit nicely to the Calabrese-Cardy formula with the same central charge $c_N =2$ (see Fig. \ref{fig:trimer_EE_orders}). It is well-known that the SU($3$)-symmetric ULS model has the SU($3$) level-$1$ conformal field theory description of its low-energy excitations \cite{itoi97}. We might speculate the same theory to govern the low-energy behavior in the trimer phase, especially at the exact SU($3$)-symmetric point $\theta=\pi/2$.

\begin{figure}[htbp]
\includegraphics*[width=0.4\textwidth]{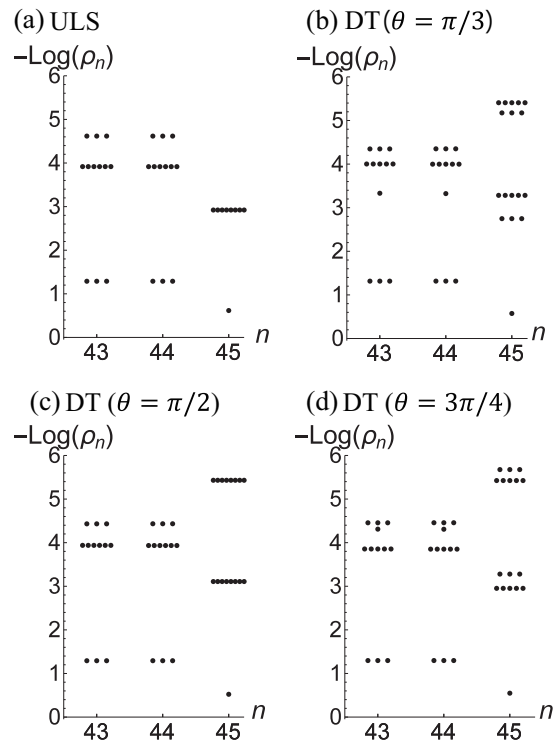}
\caption{
Entanglement spectrum for (a) ULS model \cite{PJkim16} and (b)-(d) DT model at various angles $\theta$, starting from the leftmost edge cut $n=1$. For each cut position $n$, a perfect correspondence in the degeneracy of the low-lying levels of ULS model with the pure trimer model ($\theta=\pi/2$) exists; compare (a) and (c). Degeneracies get lifted as $\theta$ deviates from $\pi/2$; see (b) and (d) plots.}
\label{fig:ES}
\end{figure}

The similarity of the pure trimer model with the ULS model, both SU($3$)-symmetric, extends into the entanglement spectrum. The degeneracy structure in the low-lying ES exhibits a perfect agreement between the two models as shown in Fig. \ref{fig:ES}. As one moves away from the pure trimer point and the symmetry of the Hamiltonian lowered, some degeneracies get lifted as shown in Fig. \ref{fig:ES}. Despite the lowered symmetry, the Calabrese-Cardy formula fit of the entanglement entropy continues to work well through the entire trimer region with the same central charge $c=2$.

The absence of energy gap in both ULS and pure trimer models are guaranteed by the Lieb--Schultz--Mattis-type theorem for SU($N$) spin models proven by Affleck and Lieb~\cite{affleck86}.
A more recent and advanced suggestion that any symmetric representation of the SU($3$) spin with a single row of Young tableau boxes of length $p$ not equal to a multiple of 3 should give a gapless spectrum\cite{greiter07,affleck17} is consistent with our observation, since both ULS and pure trimer models are written in the fundamental representation of SU(3) with $p=1$. This symmetry is presumably lowered to U(1)$\times$U(1) or SU(2)$\times$U(1) as the angle $\theta$ moves away from $\pi/2$. In both cases the overall central charge remains at 1+1=2. The gapless modes of the ULS model are the spinons carrying the fractionalized spin quantum number~\cite{sutherland75}. Guided by the identical low-level entanglement spectrum structure we may speculate that the low-energy physical excitations of the pure trimer model are also the spinons~\cite{cho17}.

\begin{figure}[htbp]
\includegraphics*[width=0.49\textwidth]{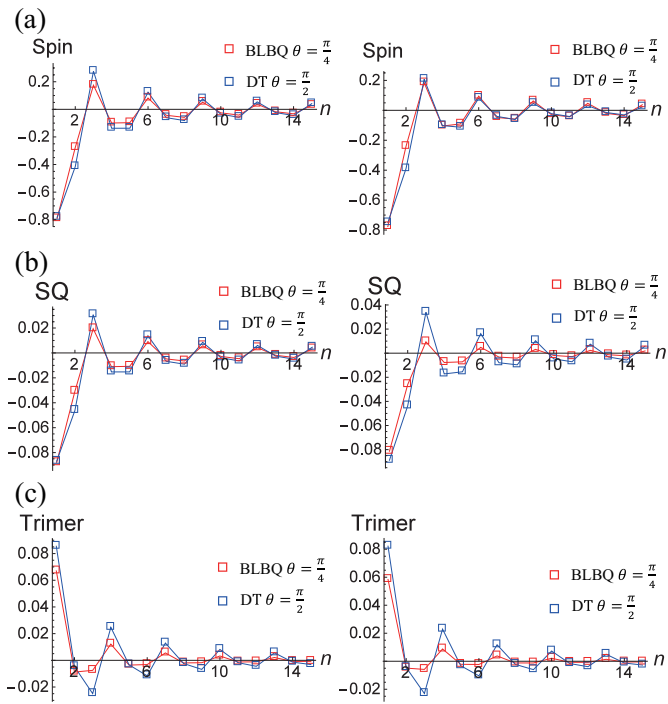}
\caption{
(a) Spin, (b) spin quadrupole, and (c) trimer correlation functions for BLBQ and DT models. Left columns show overlapping results for all correlation functions at $\theta = \pi/4$ for BLBQ and $\theta = \pi/2$ for DT model. Both models are SU(3)-symmetric. Right column shows results away from SU(3) symmetry. Overall behavior of correlation functions in the two phases are extremely similar.  }
\label{fig:trimer_BLBQ_corr}
\end{figure}

Algebraically decaying spin, spin-quadrupolar, and trimer correlation functions are shown in Fig. \ref{fig:trimer_BLBQ_corr}. For the spin-quadrupolar order we used $(S_i^z )^2$ as the operator. An interesting observation arises from our numerical investigation. For all three correlations, the ULS and the pure trimer models, both SU($3$)-symmetric, display nearly identical results. As one moves away from either pure trimer or the ULS point, behavior of the correlation functions also changes, but not by a great deal. The near-perfect identity of the correlation functions at the respective SU($3$)-symmetric points, together with their slow variation with the angle $\theta$, suggest that the trimer phase we identify in the DT model is the same phase as the spin-quadrupolar phase of the BLBQ model. A large body of the BLBQ literature refers to this phase as spin-quadrupolar, and intentionally avoids the use of the term ``trimer", which is associated with the translational symmetry breaking by three sites. Here we introduce a more carefully chosen phrase ``trimer liquid" to refer to the phase without the loss of translational symmetry, and strongly remind the readers that physical properties of the trimer liquid are identical to those of the spin-quadrupolar phase. It comes down to semantics, it seems, to choose the preferred terminology to describe the phase. Alternatively, one can refer to both the spin-quadrupolar and the trimer liquid phases as SU(3)$_1$ spin liquid phase, which would be technically the more precise characterization.

The remarkably similar behavior in the correlation functions of the two SU(3) models, ULS on one hand and pure trimer on the other, are surprising. To refresh our memory, recall that the models are given in terms of pairwise exchange operators $P_{ij}$ as
\begin{align}
H_{\rm ULS} &= \sum_{i} P_{ij} , \nn
H_{\rm PT} &= \sum_i \Bigl( 2P_{ij}  + P_{ik} - P_{ijk} - P^{-1}_{ijk} \Bigr) .
\end{align}
For lattice sizes $L = 3, 6, 9$ with periodic boundary conditions, the overlap of the exact ground states of the two models is very good, i.e., 1, 0.985654, 0.97127, respectively. A more thorough investigation of the remarkable similarity in the ground state properties of the two models are, however, delegated to future investigation.

\subsection{Macroscopically degenerate phase}
The fourth phase under consideration occurs over $\pi \le \theta \le 3\pi/2$, where both coefficients of the dimer and trimer projection operators in the DT model have non-negative values. It is flanked on either side by the pure bi-quadratic and pure trimer Hamiltonians with positive coefficients. The corresponding region in the BLBQ model is called the ferromagnetic (FM) phase, which includes the pure ferromagnetic Heisenberg exchange model. Except at the end points of the FM phase where there are known macroscopic degeneracies in the ground states, the FM ground state is unique up to the global SO(3) rotation. By contrast, as we shall see below, the entire MD phase has finite zero-temperature entropy.

The $\theta=\pi$, anti-PBQ point is known to possess macroscopically degenerate ground states whose number grows exponentially with the system size~\cite{nomura91}. The number of the ground states is bounded from below by that of the classical (direct-product) states that are annihilated by ${\cal P}_D (i)$ for all $i$. One can count such classical states using a standard transfer-matrix method (see Appendix \ref{app:transfer matrix} for details). A lower bound so obtained grows with the system size $N$ as $2^N$ for large $N$. A similar counting yields lower bounds for the number of ground states for other values of $\theta$. For $\theta=3\pi/2$ (the anti-trimer limit) and $\pi < \theta < 3\pi/2$, lower bounds are obtained as $(2.414)^N$ and $2 \times (1.618)^N$, respectively. A detailed derivation is given in Appendix \ref{app:transfer matrix}.

\begin{table*}[htb]
\begin{tabular}{c|cccccccc|c}
\hline \hline
$N$ & \,~3~\, & 4 & 5 &6 & 7 & 8 & 9 & 10 & OEIS \\ \hline
$\theta=\pi$ & ~~18 ~ & ~47~ & ~123 ~ & ~322~ & ~843 ~ & ~2207~ & ~5778~ & ~15127~ & A005248  \\
$\pi < \theta < 3\pi/2$ & \,~17~\, & ~41~ & ~83~ & ~209~ & ~479~ & ~1169~ & ~2787~ & ~6745~ & $-$  \\
$\theta=3\pi/2$ & \,~26~\, & ~72~ & ~198~ & ~570~ & ~1641~ & ~4725~ & ~13605~ & ~39174~ & $-$ \\
\hline \hline
\end{tabular}
\caption{Ground-state degeneracy of the periodic chain up to $N=10$ sites. The rightmost column shows the corresponding integer sequences in the On-Line Encyclopedia of Integer Sequences (OEIS).}
\label{tab:PBC}
\end{table*}
\begin{table*}[htb]
\begin{tabular}{c|cccccccc|c}
\hline \hline
$N$ & \,~3~\, & 4 & 5 &6 & 7 & 8 & 9 & 10 & OEIS  \\ \hline
$\theta=\pi$ & ~~21 ~ & ~55~ & ~144 ~ & ~377~ & ~987 ~ & ~2584~ & ~6765~ & ~17711~ & A001906   \\
$\pi < \theta < 3\pi/2$ & \,~20~\, & ~49~ & ~119~ & ~288~ & ~696~ & ~1681~ & ~4059~ & ~9800~ & A048739 \\
$\theta=3\pi/2$ & \,~26~\, & ~75~ & ~216~ & ~622~ & ~1791~ & ~5157~ & ~14849~ & ~42756~  &
A076264 \\
\hline \hline
\end{tabular}
\caption{Ground-state degeneracy of the open chain up to $N=10$ sites. The rightmost column shows the corresponding integer sequences in the On-Line Encyclopedia of Integer Sequences (OEIS).}
\label{tab:OBC}
\end{table*}

We have also carried out a numerical counting of the ground state degeneracy by means of exact diagonalization of the Hamiltonian for growing system sizes in both periodic and open chains.
Let $Z_N$ be the number of ground states of the chain of length $N$.
The results for $Z_N$ of periodic chains are summarized in Table \ref{tab:PBC}.
The integer sequence found at $\theta=\pi$ under the periodic boundary condition (first row in Table I) matches A005248 in the On-Line Encyclopedia of Integer Sequences (OEIS)~\cite{OEIS}, which is called the bisection of Lucas numbers~\footnote{Lucas numbers are defined by the recurrence relation, $L_N = L_{N-1}+L_{N-2}$ with the initial conditions, $L_0=2, L_1=1$.}, and is defined by the recurrence relation,

\begin{align}
Z_N = 3 Z_{N-1} - Z_{N-2}, \quad Z_1 = 3, Z_2= 7 .
\label{eq:rec1}
\end{align}
It follows from the above relation that $Z_N$ for large $N$ behaves as $Z_N \sim \varphi^{2N} \sim (2.618)^N$, where $\varphi =(1+\sqrt{5})/2$ is the golden ratio. Therefore, the true ground-state degeneracy grows much faster than the lower bound $2^N$ since it counts entangled ground states that cannot be written in the form of classical states.
The ground-state entropy per site is given by $s=\ln Z_N/N \sim 2 \ln \varphi \sim 0.962$. The other two sequences do not match anything in the OEIS. However, from the fit to the data, we find $Z_N \sim (2.412)^N$ for $\pi < \theta < 3 \pi/2$ and $Z_N \sim (2.879)^N$ for $\theta=3 \pi/2$. Again, they grow faster than the lower bounds estimated from the number of classical states.

The ground-state degeneracies in the open chains are summarized in Table \ref{tab:OBC}. The sequence for $\theta=\pi$ coincides with the bisection of Fibonacci sequence (A001906 in OEIS).
The recurrence relation for the sequence is
\begin{align}
Z_N = 3 Z_{N-1} -Z_{N-2}, \quad Z_1 = 3, Z_2=8,
\end{align}
which is the same as Eq. (\ref{eq:rec1}) with different initial conditions. Therefore, the ground-state entropy per site $s$ is also the same in the thermodynamic limit. The sequence for $\pi < \theta < 3 \pi/2$ matches with the partial sums of Pell numbers (A048739 in OEIS), defined by the recurrence relation
\begin{align}
Z_N = 3 Z_{N-1}-Z_{N-2}-Z_{N-3}, \quad Z_1 = 3, Z_2 = 8, Z_3=20,
\end{align}
from which it follows that $Z_N \sim (1+\sqrt{2})^N \sim (2.414)^N$ for large $N$. The sequence for $\theta=3 \pi/2$ matches with the number of ternary $(0,1,2)$ sequences without a consecutive $012$ (
A076264 in OEIS). It is generated by the following rule
\begin{align}
Z_N = 3 Z_{N-1}- Z_{N-3}, \quad Z_1 = 3, Z_2 = 9, Z_3 =26,
\end{align}
from which it follows that $Z_N \sim (x^*)^N$, where $x^* \sim 2.874$ is the largest root of the cubic equation $x^3+3x^2+1=0$.

To summarize, we conjecture that the exact value of the ground-state entropy per site in the thermodynamic limit is
\begin{align}
s = \left\{
\begin{array}{rclc}
2 \ln \varphi & \!\sim\! & 0.962 & \quad \theta=\pi \\
\ln (1+\sqrt{2}) & \!\sim\! & 0.881 & \quad \pi < \theta < 3\pi/2 \\
\ln x^* & \!\sim\! & 1.06 & \quad \theta=3 \pi/2
\end{array}
\right.,
\end{align}
where $\varphi=(1+\sqrt{5})/2$ and $x^* \sim 2.874$.
A rigorous proof of this conjecture is, however, beyond the scope of the present study and is left for future work.

\section{Discussion}
Motivated by the two competing tendencies in the spin-$1$ chain to form a singlet as either a dimer or a trimer, we proposed a dimer-trimer Hamiltonian with relative interaction strengths parameterized by the angle $\theta$. Dimer and trimer phases are realized respectively as one interaction becomes dominant over the other. The dimer phase is adiabatically connected to the phase of the same name in the well-known bilinear-biquadratic spin-$1$ Hamiltonian. The trimer liquid phase is gapless and critical, with central charge $c=2$, and is likely the same phase as the spin quadrupolar phase of the BLBQ model. The dimer and trimer phases are separated by the SPT phase on one side and the macroscopically degenerate phase on the other. Our work has primarily focused on carving out various phases of the dimer-trimer model. Although considerable further effort is required to clarify the nature of each phase transition, our current numerics seems to be consistent with all phase boundaries being first-order.

Various extensions of the BLBQ Hamiltonian that include the second-neighbor and three-site exchange have been studied numerically and using field-theoretic ideas in recent years \cite{pixley14,chepiga16}. These models are adiabatic extensions of the BLBQ model in the sense that turning off certain interaction parameters gives back the BLBQ Hamiltonian. The DT model, on the other hand, certainly is not such an adiabatic extension in the microscopic sense of the Hamiltonian, and yet its phase diagram turns out to be extremely similar to that of the BLBQ model. All aspects of our investigation indicate that the nature of dimer, trimer liquid, and SPT phases are identical in the two models. Significant overlaps found in the ground state wavefunctions and correlation functions of the SU(3) ULS model and the pure trimer model render further support to the deep connection bridging the two models. How to understand this connection is an interesting future challenge.
\\

\acknowledgments
We are extremely grateful to G. Y. Cho and Masaki Oshikawa for insightful discussions and comments on our work. Y.-T.O. was supported by the Global Ph.D. Fellowship Program through the National Research Foundation of Korea (NRF) funded by the Ministry of Education (No. NRF-2014H1A2A1018320).
He expresses his gratitude to Panjin Kim for providing his legacy DMRG simulation code and instructions for running it. H. K. was supported by JSPS KAKENHI Grants No. JP15K17719 and No. JP16H00985.
H.-Y. L. was supported by Ministry of Education, Culture, Sports, Science and Technology as an Exploratory Challenge on Post-K computer (Frontiers of Basic Science: Challenging the Limits).

\appendix
\section{MPO expression for trimer operator}

This appendix is devoted to the description of the matrix-product-operator (MPO) expression for the three-body trimer operator.
Compared to the MPO expressions for the two-body operators, the three-body interaction MPO is not as well-known.
We hope that our expression is helpful for future research in models involving such many-body operators.

We introduce a block representation to understand the interacting MPO's over many sites.
Indeed, with this method, one can write down the MPO for the interactions of any range.
The DT Hamiltonian is given by the contraction of blocks of MPO $A$'s,
\begin{equation}\label{DT-MPO-1}
H = \sum_{\{s_i , s_j'\}} \sum_{\{\alpha_i\}} A_{20 \alpha_1}^{s_1s_1'}A_{\alpha_2 \alpha_3}^{s_2 s_2'}
\cdots A_{\alpha_{N-1}1}^{s_Ns_N'} | s_1 \cdots\rangle \langle  s' _1 \cdots|,
\end{equation}
where all the virtual indices $\{\alpha_i\}$ run from $1$ to $20$.
The virtual indices are traced out except for the first index of $A$ at the far left of the MPO product, and the second index at the far right, as shown in the above formula with $\alpha_0=20$ and $\alpha_N = 1$.
The physical spins at the site $i$ appear as upper indices $\{s_i\}$.
The matrix representation for the MPO $A$ is given by the sum of two separate $20$ by $20$ matrices $A_1$ and $A_2$ as $A=A_1 + A_2$:

\begin{widetext}
\setcounter{MaxMatrixCols}{20}
\begin{align}
&A_1 =
\begin{bmatrix}
I                 & 0 & 0 & 0 & 0 & 0 & 0 & 0 & 0 & 0 & 0 & 0 & 0 & 0 & 0 & 0 & 0 & 0 & 0 & 0 \\
J_D \lambda_{1,1} & 0 & 0 & 0 & 0 & 0 & 0 & 0 & 0 & 0 & 0 & 0 & 0 & 0 & 0 & 0 & 0 & 0 & 0 & 0\\
J_D \lambda_{1,2} & 0 & 0 & 0 & 0 & 0 & 0 & 0 & 0 & 0 & 0 & 0 & 0 & 0 & 0 & 0 & 0 & 0 & 0 & 0\\
J_D \lambda_{1,3} & 0 & 0 & 0 & 0 & 0 & 0 & 0 & 0 & 0 & 0 & 0 & 0 & 0 & 0 & 0 & 0 & 0 & 0 & 0\\
J_D \lambda_{2,1} & 0 & 0 & 0 & 0 & 0 & 0 & 0 & 0 & 0 & 0 & 0 & 0 & 0 & 0 & 0 & 0 & 0 & 0 & 0\\
J_D \lambda_{2,2} & 0 & 0 & 0 & 0 & 0 & 0 & 0 & 0 & 0 & 0 & 0 & 0 & 0 & 0 & 0 & 0 & 0 & 0 & 0\\
J_D \lambda_{2,3} & 0 & 0 & 0 & 0 & 0 & 0 & 0 & 0 & 0 & 0 & 0 & 0 & 0 & 0 & 0 & 0 & 0 & 0 & 0\\
J_D \lambda_{3,1} & 0 & 0 & 0 & 0 & 0 & 0 & 0 & 0 & 0 & 0 & 0 & 0 & 0 & 0 & 0 & 0 & 0 & 0 & 0\\
J_D \lambda_{3,2} & 0 & 0 & 0 & 0 & 0 & 0 & 0 & 0 & 0 & 0 & 0 & 0 & 0 & 0 & 0 & 0 & 0 & 0 & 0\\
J_D \lambda_{3,3} & 0 & 0 & 0 & 0 & 0 & 0 & 0 & 0 & 0 & 0 & 0 & 0 & 0 & 0 & 0 & 0 & 0 & 0 & 0\\
0 & 0 & 0 & 0 & 0 & 0 & 0 & 0 & 0 & 0 & 0 & 0 & 0 & 0 & 0 & 0 & 0 & 0 & 0 & 0\\
0 & 0 & 0 & 0 & 0 & 0 & 0 & 0 & 0 & 0 & 0 & 0 & 0 & 0 & 0 & 0 & 0 & 0 & 0 & 0\\
0 & 0 & 0 & 0 & 0 & 0 & 0 & 0 & 0 & 0 & 0 & 0 & 0 & 0 & 0 & 0 & 0 & 0 & 0 & 0\\
0 & 0 & 0 & 0 & 0 & 0 & 0 & 0 & 0 & 0 & 0 & 0 & 0 & 0 & 0 & 0 & 0 & 0 & 0 & 0\\
0 & 0 & 0 & 0 & 0 & 0 & 0 & 0 & 0 & 0 & 0 & 0 & 0 & 0 & 0 & 0 & 0 & 0 & 0 & 0\\
0 & 0 & 0 & 0 & 0 & 0 & 0 & 0 & 0 & 0 & 0 & 0 & 0 & 0 & 0 & 0 & 0 & 0 & 0 & 0\\
0 & 0 & 0 & 0 & 0 & 0 & 0 & 0 & 0 & 0 & 0 & 0 & 0 & 0 & 0 & 0 & 0 & 0 & 0 & 0\\
0 & 0 & 0 & 0 & 0 & 0 & 0 & 0 & 0 & 0 & 0 & 0 & 0 & 0 & 0 & 0 & 0 & 0 & 0 & 0\\
0 & 0 & 0 & 0 & 0 & 0 & 0 & 0 & 0 & 0 & 0 & 0 & 0 & 0 & 0 & 0 & 0 & 0 & 0 & 0\\
0 & \lambda_{1,1} & \lambda_{1,2} & \lambda_{1,3} & \lambda_{2,1} & \lambda_{2,2} & \lambda_{2,3} & \lambda_{3,1} & \lambda_{3,2} & \lambda_{3,3} & 0 & 0 & 0 & 0 & 0 & 0 & 0 & 0 & 0 & I
\end{bmatrix},
\nn
&A_2=\nn
&
\begin{bmatrix}
0 & 0 & 0 & 0 & 0 & 0 & 0 & 0 & 0 & 0 & 0 & 0 & 0 & 0 & 0 & 0 & 0 & 0 & 0 & 0 \\
0 & 0 & 0 & 0 & 0 & 0 & 0 & 0 & 0 & 0 & 0 & 0 & 0 & 0 & J_T \lambda_{3,3} & -J_T \lambda_{3,2} & 0 & -J_T \lambda_{2,3} & J_T \lambda_{2,2} & 0\\
0 & 0 & 0 & 0 & 0 & 0 & 0 & 0 & 0 & 0 & 0 & 0 & 0 & -J_T \lambda_{3,3} & 0 & J_T \lambda_{3,1} & J_T \lambda_{2,3} & 0 & -J_T \lambda_{2,1} & 0\\
0 & 0 & 0 & 0 & 0 & 0 & 0 & 0 & 0 & 0 & 0 & 0 & 0 & J_T \lambda_{3,2} & -J_T \lambda_{3,1} & 0 & -J_T \lambda_{2,2} & J_T \lambda_{2,1} & 0 & 0\\
0 & 0 & 0 & 0 & 0 & 0 & 0 & 0 & 0 & 0 & 0 & -J_T \lambda_{3,3} & J_T \lambda_{3,2} & 0 & 0 & 0 & 0 & J_T \lambda_{1,3} & -J_T \lambda_{1,2} & 0\\
0 & 0 & 0 & 0 & 0 & 0 & 0 & 0 & 0 & 0 & 0 & -J_T \lambda_{3,3} & J_T \lambda_{3,2} & 0 & 0 & 0 & 0 & J_T \lambda_{1,3} & -J_T \lambda_{1,2} & 0\\
0 & 0 & 0 & 0 & 0 & 0 & 0 & 0 & 0 & 0 & -J_T \lambda_{3,2} & J_T \lambda_{3,1} & 0 & 0 & 0 & 0 & 0 & -J_T \lambda_{1,2} & J_T \lambda_{1,1} & 0\\
0 & 0 & 0 & 0 & 0 & 0 & 0 & 0 & 0 & 0 & 0 & J_T \lambda_{2,3} & -J_T \lambda_{2,2} & 0 & -J_T \lambda_{1,3} & J_T \lambda_{1,2} & 0 & 0 & 0 & 0\\
0 & 0 & 0 & 0 & 0 & 0 & 0 & 0 & 0 & 0 & -J_T \lambda_{2,3} & 0 & J_T \lambda_{2,1} & J_T \lambda_{1,3} & 0 & -J_T \lambda_{1,1} & 0 & 0 & 0 & 0\\
0 & 0 & 0 & 0 & 0 & 0 & 0 & 0 & 0 & 0 & J_T \lambda_{2,2} & -J_T \lambda_{2,1} & 0 & -J_T \lambda_{1,2} & J_T \lambda_{1,1} & 0 & 0 & 0 & 0 & 0\\
\lambda_{1,1} & 0 & 0 & 0 & 0 & 0 & 0 & 0 & 0 & 0 & 0 & 0 & 0 & 0 & 0 & 0 & 0 & 0 & 0 & 0\\
\lambda_{1,2} & 0 & 0 & 0 & 0 & 0 & 0 & 0 & 0 & 0 & 0 & 0 & 0 & 0 & 0 & 0 & 0 & 0 & 0 & 0\\
\lambda_{1,3} & 0 & 0 & 0 & 0 & 0 & 0 & 0 & 0 & 0 & 0 & 0 & 0 & 0 & 0 & 0 & 0 & 0 & 0 & 0\\
\lambda_{2,1} & 0 & 0 & 0 & 0 & 0 & 0 & 0 & 0 & 0 & 0 & 0 & 0 & 0 & 0 & 0 & 0 & 0 & 0 & 0\\
\lambda_{2,2} & 0 & 0 & 0 & 0 & 0 & 0 & 0 & 0 & 0 & 0 & 0 & 0 & 0 & 0 & 0 & 0 & 0 & 0 & 0\\
\lambda_{2,3} & 0 & 0 & 0 & 0 & 0 & 0 & 0 & 0 & 0 & 0 & 0 & 0 & 0 & 0 & 0 & 0 & 0 & 0 & 0\\
\lambda_{3,1} & 0 & 0 & 0 & 0 & 0 & 0 & 0 & 0 & 0 & 0 & 0 & 0 & 0 & 0 & 0 & 0 & 0 & 0 & 0\\
\lambda_{3,2} & 0 & 0 & 0 & 0 & 0 & 0 & 0 & 0 & 0 & 0 & 0 & 0 & 0 & 0 & 0 & 0 & 0 & 0 & 0\\
\lambda_{3,3} & 0 & 0 & 0 & 0 & 0 & 0 & 0 & 0 & 0 & 0 & 0 & 0 & 0 & 0 & 0 & 0 & 0 & 0 & 0\\
0 & 0 & 0 & 0 & 0 & 0 & 0 & 0 & 0 & 0 & 0 & 0 & 0 & 0 & 0 & 0 & 0 & 0 & 0 & 0
\end{bmatrix}
.
\end{align}
\end{widetext}
Here, $J_D=-\cos\theta$ and $J_T=-\sin\theta$ are coefficients of the DT model (\ref{eq:DT_model}). All elements of the above matrices are themselves $3$ by $3$ matrices, whose indices correspond to the physical spin $s_i$ and $s_i'$ at the local site $i$, with the identity matrix $I$ and $\lambda_{\alpha,\beta}^{s_i s_i'}= \delta_{\alpha s_i}\delta_{\beta s_i'}$.
The representation of the spin operators $S_i^{jk} = -i \varepsilon_{ijk}$ is employed, where $\varepsilon_{ijk}$ is the anti-symmetric tensor. In this representation, the dimer operator $D$ and the trimer operator $T$ are given by
\begin{align} \label{eq:d-t-spin-rep}
D(n) = & \sum_{x,y} \lambda_{x,y} (n) \lambda_{x,y} (n+1) \nn
T(n) = & \sum_{x,y,z,l,m,n} \varepsilon_{xyz}\varepsilon_{lmn}~ \lambda_{x,l} (n)\lambda_{y,m} (n+1 )\lambda_{z,n} (n+2) ,\nn
\end{align}
where the argument of $\lambda$ represents the local site index $n$.
One can check that the product of $A_1$ in the matrix representation of $A$ over the whole lattice yields the pure-biquadratic model. In order to generate the trimer interaction, we must add the $A_2$ term in the MPO $A$, and multiply $A = A_1 + A_2$ through the entire lattice.

\begin{figure*}[htbp]
\includegraphics*[width=0.9\textwidth]{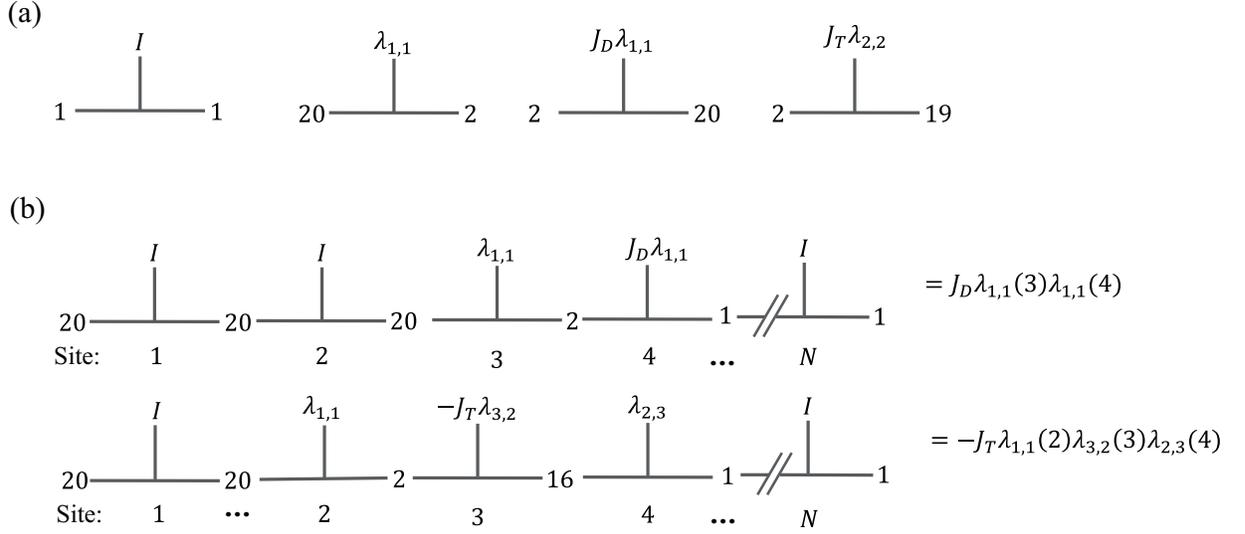}
\caption{
(a) Examples of the comb representation for the nonzero elements in the MPO matrix $A$. The left and right ends of each graph correspond to the virtual indices of $A$. The ``tooth" of each graph represents the physical operators such as $I$ and $\lambda_{\alpha, \beta}$.
(b) Examples of contracted graphs representing a specific term in the Hamiltonian. The connecting links of the adjacent blocks must have the same virtual index. In the first low, the resulting product is $J_D \lambda_{1,1} (3) \lambda_{2,2} (4)$, which is one of the terms in the dimer operator at site 3, i.e., $D(3)$ in (\ref{eq:d-t-spin-rep}). The argument of $\lambda$ denotes the spatial position. In the second row, the product gives $-J_T \lambda_{1,1} (2) \lambda_{3,2} (3)\lambda_{2,3} (4)$, which is one of the terms in the trimer operator $T(2)$ defined in (\ref{eq:d-t-spin-rep}).}
\label{fig:mpo_comb}
\end{figure*}

It is instructive to work with the ``comb representation" to guide one's understanding of the structure of $A$. In Fig. \ref{fig:mpo_comb}(a), we show how to make a graphical representation of the nonzero elements in the matrix $A$. By contracting the graphs with the same virtual indices as shown in Fig. \ref{fig:mpo_comb}(b), we can obtain the comb-shaped graph representation for all the terms in the DT Hamiltonian.

\section{Classical states in MD phase}
\label{app:transfer matrix}
In this appendix, we present a transfer-matrix analysis for the counting of classical ground states annihilated by dimer/trimer projection operators. At $\theta=\pi$, any state which is annihilated by all dimer projection operators ${\cal P}_D (i)$ ($i=1,...,N$) is a zero-energy ground state of the Hamiltonian. Such states are not hard to construct. In the basis in which $S^z$ is diagonal, ${\cal P}_D (i)$ annihilates the state if spins at sites $i$ and $i+1$ do not add up to zero:
\begin{equation}
{\cal P}_D (i) | m_i, m_{i+1} \rangle = 0 ~ {\rm if} ~  m_i + m_{i+1} \neq 0, \label{eq:dimer-condition}
\end{equation}
where $m_i = +1, 0$ or $-1$ is an eigenvalue of $S^z_i$. Therefore, a product state which does not contain any of the configurations $00$, $+-$, and $-+$ in any neighboring sites is a ground state at $\theta=\pi$. It is then straightforward to construct a transfer matrix and count the number of such classical states exactly. The number of classical states in the periodic chain of length $N$ is obtained as $Z^{({\rm cl})}_N={\rm Tr}[T^N]$, where
\begin{align}
T = \begin{pmatrix}
1 & 1 & 0 \\
1 & 0 & 1 \\
0 & 1 & 1
\end{pmatrix}
\end{align}
with the order of the basis states $\{ |+\rangle, |0\rangle, |-\rangle \}$. Since the eigenvalues of $T$ are $\{ 2, 1,-1 \}$, we have $Z^{({\rm cl})}_N = 2^N+1+(-1)^N$.

Let us next consider the anti-trimer limit $\theta=3\pi/2$. At this point, any state which is annihilated by all trimer projections ${\cal P}_T (i)$ ($i=1,...,N$) is a zero-energy ground state of the Hamiltonian. A single trimer wavefunction (\ref{eq:trimer-wf}) has all the basis states different from one another over the three consecutive sites. Therefore, a product state which does not contain a permutation of $+0-$ in any three consecutive sites is a ground state at $\theta=3\pi/2$. One can again count the number of such states by constructing a transfer matrix:
\begin{equation}
T = \begin{pmatrix}
1 & 1 & 1 & 0 & 0 & 0 & 0 & 0 & 0 \\
0 & 0 & 0 & 1 & 1 & 0 & 0 & 0 & 0 \\
0 & 0 & 0 & 0 & 0 & 0 & 1 & 0 & 1 \\
1 & 1 & 0 & 0 & 0 & 0 & 0 & 0 & 0 \\
0 & 0 & 0 & 1 & 1 & 1 & 0 & 0 & 0 \\
0 & 0 & 0 & 0 & 0 & 0 & 0 & 1 & 1 \\
1 & 0 & 1 & 0 & 0 & 0 & 0 & 0 & 0 \\
0 & 0 & 0 & 0 & 1 & 1 & 0 & 0 & 0 \\
0 & 0 & 0 & 0 & 0 & 0 & 1 & 1 & 1
\end{pmatrix},
\end{equation}
where the order of the basis states is $\{ |++\rangle, |+0\rangle, |+-\rangle, |0+\rangle, |00\rangle, |0-\rangle, |-+\rangle, |-0\rangle, |--\rangle \}$. The largest eigenvalue of $T$, in modulus, can be obtained analytically, and is given by $\lambda_{\rm max} = 1+\sqrt{2}$. Therefore, we have $Z^{({\rm cl})}_N  \sim (\lambda_{\rm max})^N \sim (2.414)^N$ for large $N$.

Finally, we derive a lower bound for the number of ground states for $\pi < \theta < 3\pi/2$. In this region, a ground state must be annihilated by both ${\cal P}_D (i)$ and ${\cal P}_T (i)$ for all $i=1,...,N$. This happens when the spin configuration of any three consecutive sites is one of the following: $\{ |+++\rangle, |++0\rangle, |+0+\rangle, |0++\rangle, |0+0\rangle, |0-0\rangle, |0--\rangle, |-0-\rangle, |--0\rangle, |---\rangle \}$. This $3$-site rule determines the transfer matrix
\begin{equation}
T = \begin{pmatrix}
1 & 1 & 0 & 0 & 0 & 0 \\
0 & 0 & 1 & 0 & 0 & 0 \\
1 & 1 & 0 & 0 & 0 & 0 \\
0 & 0 & 0 & 0 & 1 & 1 \\
0 & 0 & 0 & 1 & 0 & 0 \\
0 & 0 & 0 & 0 & 1 & 1
\end{pmatrix},
\end{equation}
where the order of the basis states is $\{ |++\rangle, |+0\rangle, |0+\rangle, |0-\rangle, |-0\rangle, |--\rangle \}$. The eigenvalues of $T$ are given by
\begin{equation}
\lambda= \frac{1\pm \sqrt{5}}{2}, 0,
\end{equation}
each of which is two-fold degenerate. Therefore, the number of classical states in the periodic chain is obtained as
\begin{equation}
Z^{({\rm cl})}_N = {\rm Tr} [T^N]
= 2 \varphi^N + 2 (-1)^N \varphi^{-N},
\end{equation}
where $\varphi=(1+\sqrt{5})/2$ is the golden ratio. The sequence of $Z^{({\rm cl})}_N$ coincides with the Fibonacci sequence starting from $2$ and $6$ (A022112 in OEIS). For large $N$, we have $Z^{({\rm cl})}_N \sim 2 \varphi^N \sim 2\times (1.618)^N$.

\bibliographystyle{apsrev}
\bibliography{reference}

\end{document}